\documentclass[10pt]{article}
\usepackage[english]{babel}
\usepackage[pdftex]{graphicx}
\usepackage{color}
\definecolor{darkgreen}{rgb}{0,.5,0}
\usepackage[colorlinks,urlcolor=blue]{hyperref}

\title{The structure and modeling results of the parallel spatial switching
system}

\author{Denis Kutuzov (\url{d_kutuzov@mail.ru}), Alexey Osovsky}
\date{April 20, 2007}

\begin{document}
\maketitle
\thanks{Information Systems Dept., Astrakhan State University, 20a,
Tatischeva Str., Astrakhan, 414056, Russia,
(\url{http://www.aspu.ru/})}

\begin{abstract}
Problems of the switching parallel system designing provided spatial
switching of packets from random time are discussed. Results of
modeling of switching system as systems of mass service are
resulted.

Index Terms: parallel switching, information systems, system of mass
service, packet switching, communication network.
\end{abstract}

The increase of rate of switching system elements, as well as other
devices occurs basically due to advantages in technological base. At
the same time another decision is possible. It is perfection of
algorithms and methods of switching and creation a new switching
decisions, not dependent on technological base.

Such decision for switching systems may become a parallel process
method which would allow to increase transfer rate and to decrease
losses at processing packets.

\begin{figure}[htbp]
\centerline{\includegraphics[width=7cm]{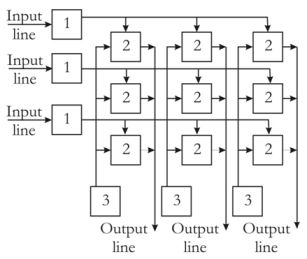}} \caption{The
structure of the parallel spatial switching system} \label{fig1}
\end{figure}

The parallel switching system \cite{Kutuzov} provides a parallel
spatial switching of the information packets at the random moments
of time and consists of the following blocks (see Fig.~\ref{fig1}):
block of synchronization of the input packet with the moment of
identification (1), elements of switching (2) and blocks of tag
generation for the target lines (3).

Operation of system is carried out with the following algorithm: a
requirement of connection establishment is transferred on the system
input. Each requirement represents the information packet having
headline (tag) which identifies a target line of the switching
system which it is necessary to send the given packet. As
requirements of the establishment connections (packets) arrived on
the system inputs at the random time, that is asynchronously, and
identification v is possible only at the certain moments of time,
the requirement (packet) should be delayed up to the moment when the
tag comparison and to start switching is possible. Such delay is
carried out by blocks of packet synchronization with the moment of
identification and may be various for the packets which have arrived
at die different time. Some requirements (packets), expecting the
beginning of identification represent a pack.

Identification represents process of bit-by-bit tag comparison for
connection lines (it contains in a packet), and the target line tags
which generated blocks 3 (Fig.~\ref{fig1}). At coincidence of the
target line tag with heading tag in the packet, the appropriate
switching element establishes connection for transfer of packet on
the system output. After transfer of the packet the connection
should be destroyed. All functions of the conflict resolution, for
example, at attempt of two and more input lines to establish
connection with one target line are assigned to switching elements 2
(Fig.~\ref{fig1}).

\begin{figure}[htbp]
\centerline{\includegraphics[width=7cm]{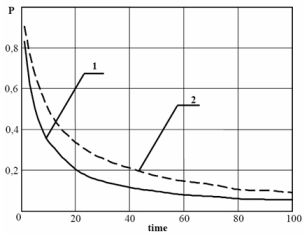}}\caption{The modeling
results of the parallel spatial switching system} \label{fig2}
\end{figure}

Diagrams of loss probabilities of packets (P) from average
(modeling) time between packet receiving are shown in
Fig.~\ref{fig2}.

It is possible to make a conclusion on reduction of loss probability
in comparison with the system with consecutive switching by results
of modeling of the parallel switching system \cite{Lunev}. The
diagram 1 in Fig.~\ref{fig2} corresponds to parallel switching
system, the diagram 2 is consecutive one.


\begin{thebibliography}{9}
\bibitem{Kutuzov} D.V.~Kutuzov, ``Parallel adjustment of matrix
switching system at dynamical requirements'' // Engineering and
technology, 2005. No. 2(8), pp. 57--59.

\bibitem{Lunev} A.P.~Lunev, I.Yu.~Petrova, D.V.~Kutuzov,
A.V.~Osovsky, ``Simulation of matrix switchboards''. The patent
regisrtation No. 2005611003, the legal owner Astrakhan State
University.

\end{thebibliography}
\end{document}